\documentclass[sigconf]{acmart}

\usepackage{enumitem}
\usepackage{algorithm}
\usepackage{algorithmic}
\setlist[itemize]{nosep, leftmargin=*}
\setlist[enumerate]{nosep, leftmargin=*}

\copyrightyear{2026}
\acmYear{2026}
\setcopyright{cc}
\setcctype{by}
\acmConference[RecSys '26]{20th ACM Conference on Recommender Systems}{September 27--October 02, 2026}{Minneapolis, MN, USA}
\acmBooktitle{20th ACM Conference on Recommender Systems (RecSys '26), September 27--October 02, 2026, Minneapolis, MN, USA}
\acmDOI{10.1145/3773078.3831771}
\acmISBN{979-8-4007-2284-4/2026/09}

\begin{document}

\title{Zero-Observation User Reactivation with Gap-Driven Dimensional Gating}

\author{Jiandong Ding}
\authornote{Corresponding author.}
\orcid{0000-0001-8123-9889}
\affiliation{%
  \institution{Fudan University}
  \city{Shanghai}
  \country{China}}
\email{jdding@fudan.edu.cn}

\author{Tianying Liu}
\orcid{0000-0003-3155-7832}
\affiliation{%
  \institution{Huawei Technologies}
  \city{Shanghai}
  \country{China}}
\email{liutianying2@huawei.com}

\author{Fuyuan Liu}
\orcid{0009-0001-5008-0876}
\affiliation{%
  \institution{Huawei Technologies}
  \city{Shanghai}
  \country{China}}
\email{liufuyuan2@huawei.com}

\author{Huijie Qin}
\orcid{0009-0005-7722-850X}
\affiliation{%
  \institution{Huawei Technologies}
  \city{Shanghai}
  \country{China}}
\email{qinhuijie@huawei.com}

\author{Tiandeng Wu}
\orcid{0009-0001-0951-0790}
\affiliation{%
  \institution{Huawei Technologies}
  \city{Shanghai}
  \country{China}}
\email{wutiandeng1@huawei.com}

\begin{abstract}
Sequential recommendation (SR) models capture continuously observed behavior, but a returning user may have no interactions for months or years. We define this setting as \textit{Zero-Observation Reactivation}: the user has a pre-gap history, while the platform observes no behavioral signals during a macro-gap $\Delta t$. Under a chronologically aligned Gap-Synthesize Protocol on three Amazon datasets (Video Games, CDs \& Vinyl, and Movies \& TV), Hit@10 decreases monotonically across the evaluated gap buckets and reaches its lowest level beyond one year. The pattern appears across recurrent, unidirectional, and bidirectional SR backbones.
 
We propose \textbf{DeltaGate}, a lightweight output-layer plugin that keeps the backbone frozen and routes each representation dimension between the personalized history and a learned, zero-initialized global prior. The gate is conditioned jointly on $\Delta t$ and the personalized representation. In a controlled diagnostic, we hold the personalized representation fixed and vary $\Delta t$ to isolate the trained gate's response to the gap input. In the $>$365d Video Games bucket, DG-SASRec reaches 0.047 Hit@10 versus 0.031 for SASRec, while DG-BERT4Rec reaches 0.046 versus 0.025 for BERT4Rec, with 66K trainable parameters (2--4\% overhead). End-to-end retraining attains higher absolute accuracy but changes the backbone embeddings; the frozen plugin preserves zero backbone drift, uses about 40$\times$ fewer trainable parameters, and retains observable dimension-wise routing. The source code is available at \url{https://github.com/jdding/DeltaGate}.
\end{abstract}



\maketitle

\section{Introduction}

Sequential recommendation systems model users from ordered interaction histories~\cite{kang2018self, sun2019bert4rec, wang2021survey}. Most evaluations, however, treat the final observed state as current regardless of how long the user has been absent. Temporal interaction studies such as JODIE~\cite{kumar2019predicting} show that latent states evolve over time. After a long period without observations, the reliability of a pre-gap representation is therefore uncertain.

\begin{figure}[t]
\centering
\includegraphics[width=\linewidth]{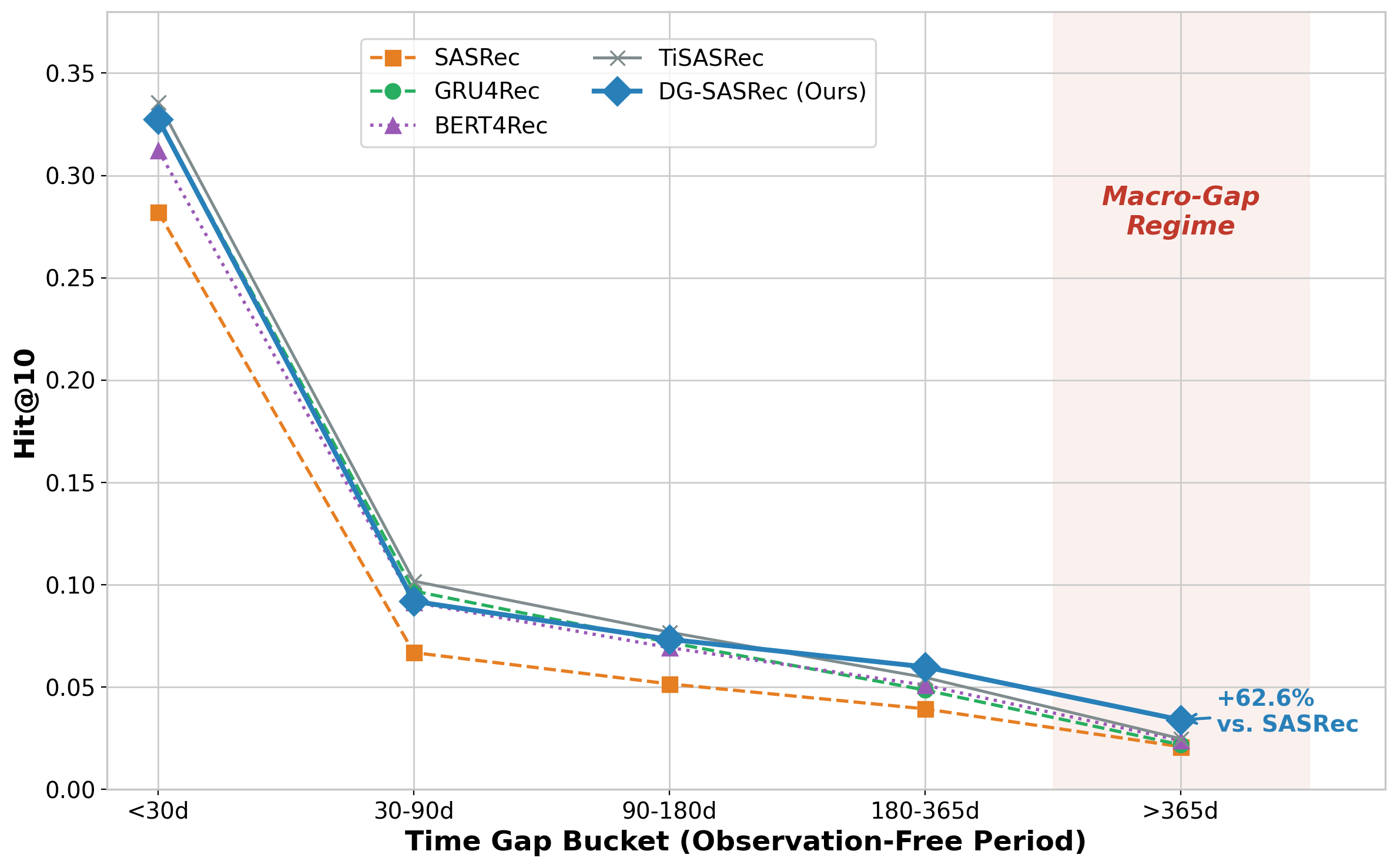}
\caption{Hit@10 across gap buckets on Movies \& TV. All evaluated baselines decline as the gap widens, while DG-SASRec recovers part of the loss beyond 365 days.}
\Description{Line plots of Hit at 10 over five increasing inactivity-gap buckets for the evaluated sequential backbones and DG-SASRec on Movies and TV. The baseline curves decline with wider gaps, and DG-SASRec is highest in the final bucket.}
\label{fig:decay}
\end{figure}

Consider a user who is absent from a platform for over a year and then returns. Unlike traditional new-user cold-start scenarios~\cite{briand2021semi, zhang2023coldwarm}, returning users may have substantial pre-gap histories. Yet, unlike continuously active users, they generate no observations during the gap. We call this setting the Zero-Observation Reactivation problem.

We evaluate this setting on three Amazon categories~\cite{ni2019justifying} using user-level chronological leave-one-out targets and natural-gap stratification. Across the tested buckets, performance declines monotonically and is lowest beyond 365 days (Figure~\ref{fig:decay}). This pattern occurs for GRU4Rec, SASRec, and BERT4Rec. TiSASRec~\cite{li2020time}, which models time intervals within the observed sequence, provides only limited recovery in this macro-gap regime.

\begin{figure}[t]
\centering
\includegraphics[width=\linewidth]{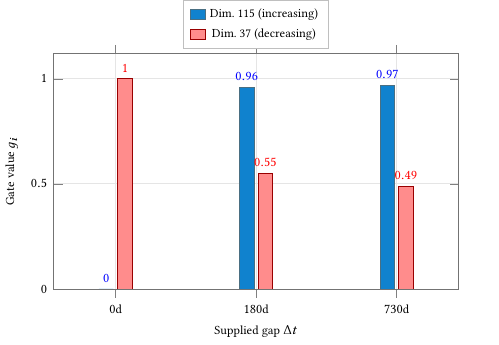}
\caption{Controlled gate diagnostic with a fixed user-history representation and varied $\Delta t$ (DG-BERT4Rec, Movies \& TV, seed 42). From 0 to 730 days, Dimension 115's history weight increases from 0.00 to 0.97, while Dimension 37's decreases from 1.00 to 0.49.}
\Description{Grouped bars at supplied gaps of 0, 180, and 730 days while one user-history representation is held fixed. Dimension 115 increases from 0.00 to 0.97, and Dimension 37 decreases from 1.00 to 0.49. Because the gate multiplies the personalized representation, higher values assign more weight to personal history.}
\label{fig:causal_probe}
\end{figure}

DeltaGate rests on the hypothesis that representation dimensions may encode interests with different temporal stability. The diagnostic in Figure~\ref{fig:causal_probe} holds the history representation fixed and varies only $\Delta t$. The resulting trajectories show a dimension-specific response to the gap input. This intervention diagnoses the trained gate's input sensitivity; it is neither a training objective nor an estimate of a population-level causal effect.

DeltaGate operationalizes this hypothesis through gap-driven routing at the representation output layer. It computes an element-wise gate from $\Delta t$ and the personalized history, then interpolates between that history and a learned global prior initialized at zero.

Our contributions are summarized as follows:
\begin{itemize}
    \item We define \textit{Zero-Observation Reactivation} and document monotonic performance decay across gap buckets for three SR backbones on three Amazon categories.
    \item We introduce a frozen-backbone, dimension-wise gate conditioned on both the inactivity gap and the user's pre-gap representation; routing-component controls compare the learned route with its personalized-history and shared-prior endpoints.
    \item We compare frozen-plugin and end-to-end adaptation. Full retraining yields the highest accuracy, while the plugin retains zero backbone drift, higher parameter efficiency, and interpretable gate behavior.
\end{itemize}

\section{Related Work}

\subsection{Sequential Recommendation}
Sequential recommendation predicts the next item from an ordered history. Representative encoders include Markov-chain models~\cite{rendle2010fpmc}, recurrent models such as GRU4Rec~\cite{hidasi2015session}, convolutional models such as Caser~\cite{tang2018caser}, graph models such as SR-GNN~\cite{wu2019srgnn}, and self-attention models such as SASRec~\cite{kang2018self}. Self-supervised objectives further improve sequence representations~\cite{zhou2020s3rec}. These methods encode the observed sequence but do not explicitly recalibrate the final user state according to a subsequent observation-free gap.

\subsection{Time-Aware Sequential Recommendation}
Time-LSTM~\cite{zhu2017what} and MEANTIME~\cite{cho2020learning} incorporate intervals into recurrent or attention-based sequence encoders. TiSASRec~\cite{li2020time} models absolute and relative intervals between observed items, while TGSRec~\cite{fan2021tgsrec} extends temporal modeling to continuous-time graphs. RPE4Rec~\cite{cheng2026rpe4rec} uses relative-position signals to capture short-term co-occurrence for dynamic item retrieval. These methods model timing within the observed interaction process. DeltaGate instead recalibrates the final representation using the observation-free gap after the last recorded interaction, without modifying the encoder internals.

\subsection{Long-term Interest vs. Long-gap Reactivation}
Lifelong models such as SIM~\cite{pi2020sim}, TWIN~\cite{chang2023twin}, Mamba4Rec~\cite{mamba4rec2024}, and hierarchical generative recommenders~\cite{chen2026flat} increase the amount or structure of observed history that a model can use. Their focus is representation learning within long observed sequences. Zero-Observation Reactivation instead asks how strongly to retain the resulting representation after no new behavior has been observed for an extended period. DeltaGate operates at this output-representation level and is evaluated with three backbone families (Section~\ref{sec:experiments}).

\subsection{User Reactivation and Cold-Start}
Cold-start recommendation typically addresses new users with little or no history~\cite{briand2021semi, zhang2023coldwarm}. Reactivated users differ because a substantial pre-gap history exists, but its current reliability is unknown. LLM-based recommenders provide richer semantic representations~\cite{bao2023tallrec, wu2024llm4rec}. DeltaGate's output-level interface is compatible in principle with systems that expose a final dense user representation for scoring, but this setting is not evaluated here. TASIF~\cite{luo2025tasif} uses frequency-aware fusion within observed sequences, whereas DeltaGate conditions on the gap after the last observation.

\subsection{Popularity Debiasing and Causal Inference}
Routing uncertain dimensions toward a global prior is related to popularity-bias research. Prior work separates user interest from conformity~\cite{zheng2021dice} or uses causal adjustment and counterfactual reasoning to reduce popularity effects~\cite{pearl2009causality, zhang2021pda, wei2021model}. DeltaGate does not estimate a causal recommendation effect or remove popularity bias. It uses the learned global prior as one endpoint of a dimension-wise interpolation and examines the resulting personalization--coverage trade-off in RQ5.

\begin{figure*}[t]
\centering
\includegraphics[width=0.96\textwidth,height=0.40\textwidth]{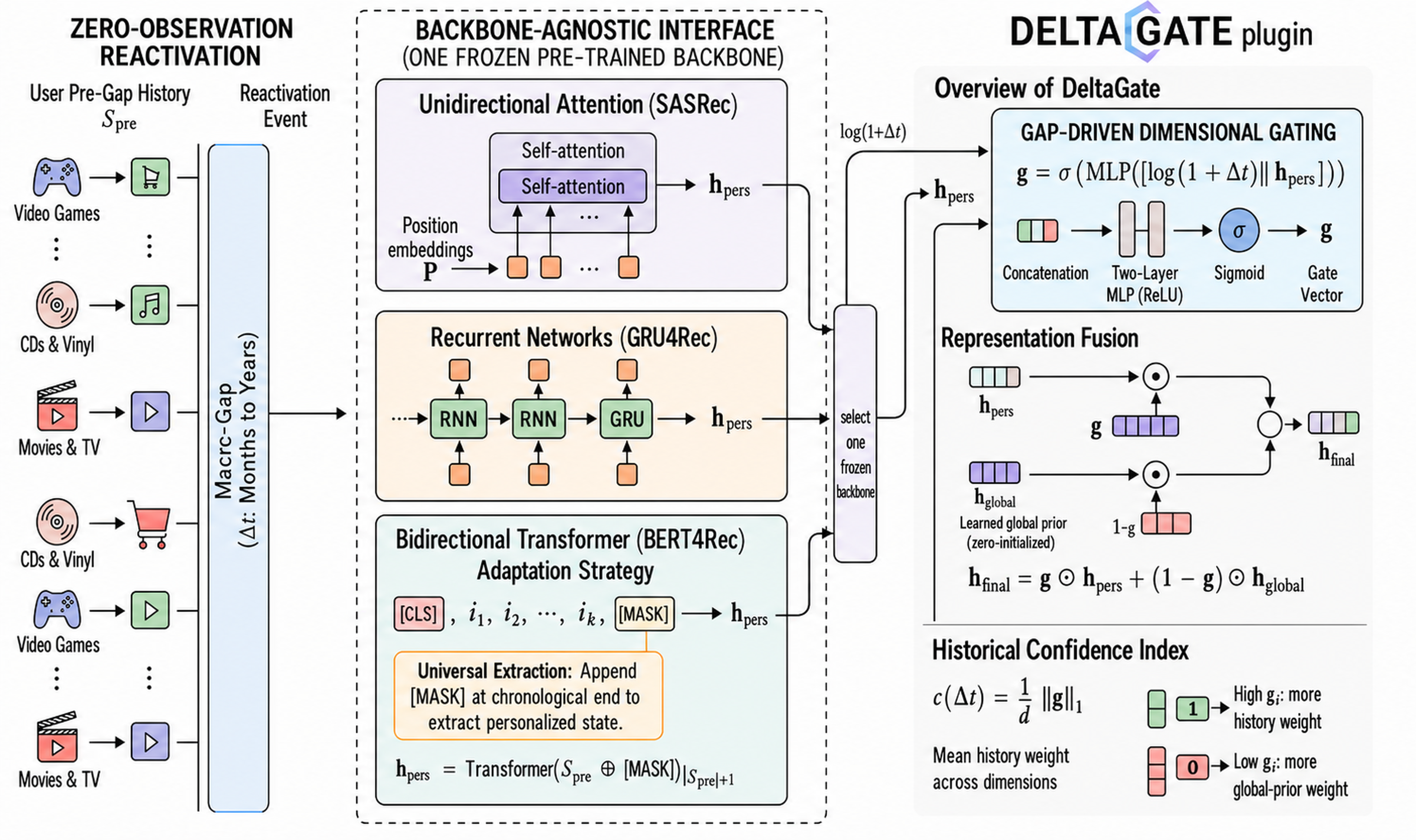}
\caption{DeltaGate architecture. One selected frozen backbone produces $\mathbf{h}_{\mathrm{pers}}$. The personalized state and the independently observed gap enter a trainable dimension-wise gate, whose output fuses $\mathbf{h}_{\mathrm{pers}}$ with a learned global prior initialized at zero. BERT4Rec exposes its final user state through an appended \texttt{[MASK]} token.}
\Description{A left-to-right architecture diagram. A pre-gap interaction sequence passes through one selected frozen backbone, either SASRec, GRU4Rec, or BERT4Rec, to produce a personalized state. The personalized state and an independently observed log-transformed gap enter a trainable dimensional gate. Two parallel branches weight the personalized state and a learned, zero-initialized global prior, then add them to form the final state. The lower panel defines the Historical Confidence Index as the mean history weight across dimensions.}
\label{fig:architecture}
\end{figure*}

\section{Problem Formulation and Protocol}
\label{sec:setup}

\subsection{The Scale and Severity of Macro-Gaps}
\label{sec:motivation}

We first measure long-gap prevalence in Amazon Video Games after 5-core filtering. Table~\ref{tab:gap_distribution} reports the test users in each $\Delta t$ bucket. This distribution motivates the macro-gap cohort studied below, but the estimate is specific to this dataset and should not be generalized to other platforms.

\begin{table}[htbp]
\centering
\caption{Distribution of test users across temporal gap buckets on Amazon Video Games.}
\label{tab:gap_distribution}
\begin{tabular}{lrr}
\toprule
$\Delta t$ Bucket & Test Users & Proportion \\
\midrule
$<$30d (Active) & 28,245 & 51.1\% \\
30--90d & 5,147 & 9.3\% \\
90--180d & 4,875 & 8.8\% \\
180--365d & 6,357 & 11.5\% \\
$>$365d (Macro-Gap) & 10,596 & 19.2\% \\
\bottomrule
\end{tabular}
\end{table}

The $>$365d cohort contains \textbf{10,596 users}, or \textbf{19.2\%} of the Video Games test users, making it a substantial macro-gap slice in this dataset.

As a non-personalized reference, Time-Decay Popularity (TDP) ranks items by recency-weighted global interaction frequency. Its Hit@10 rounds to \textbf{0.001} in the $>$365d bucket. The uniform-random reference, $10/17{,}408$, also rounds to 0.001; both remain well below the sequential baselines in Table~\ref{tab:main_results}.

\subsection{The Zero-Observation Reactivation Problem}
Let $\mathcal{U}$ and $\mathcal{I}$ denote the sets of users and items, respectively. For a user $u \in \mathcal{U}$, their chronologically ordered historical interaction sequence is $\mathcal{S}_{pre} = ((i_1,t_1),\ldots,(i_k,t_k))$, where $i_r \in \mathcal{I}$ and $t_r$ are the item and timestamp of interaction $r$. The user returns to interact with a target item $i_{target}$ at time $t_{target}$.

Standard sequential recommenders estimate the next-item probability from $\mathcal{S}_{pre}$ without conditioning the final representation on the subsequent inactivity duration. The observation-free gap is $\Delta t=t_{target}-t_k$. We study the gap-conditioned prediction problem:
\begin{equation}
    P(i_{target} \mid \mathcal{S}_{pre}, \Delta t), \qquad \Delta t \geq 0
\end{equation}
No interactions from user $u$ are observed within this interval. At prediction time, $\Delta t$ is the elapsed time since the last recorded interaction, observed when the user initiates the return session; the target item remains unknown. The model must therefore combine a potentially stale pre-gap representation with the known gap duration when predicting $i_{target}$.

\subsection{The Gap-Synthesize Evaluation Protocol}
Gap cohorts can differ in both user history and calendar-time effects such as catalog or popularity drift. Our Gap-Synthesize Protocol constructs gap-conditioned evaluation cohorts from naturally observed terminal gaps; it does not alter timestamps or synthesize user interactions. The protocol uses user-level chronological targets and natural-gap stratification to make the gap comparison consistent while avoiding a single global time cutoff:
\begin{itemize}
    \item \textbf{User-level leave-one-out.} For each user, the chronologically last interaction is the target $i_{target}$ and preceding interactions form $\mathcal{S}_{pre}$.
    \item \textbf{Natural-gap identification and masking.} The natural interval before $i_{target}$ defines $\Delta t$. Interactions unavailable at prediction time are excluded, and users must satisfy $|\mathcal{S}_{pre}|\geq5$.
    \item \textbf{Bucket stratification.} Users are grouped by $\Delta t$: \texttt{<30d}, \texttt{30-90d}, \texttt{90-180d}, \texttt{180-365d}, and \texttt{>365d}.
\end{itemize}

All models are evaluated by full ranking over the item vocabulary $\mathcal{I}$. This avoids the sensitivity of sampled metrics to the negative-sampling protocol~\cite{krichene2020sampled}. The protocol does not remove catalog drift within long natural gaps; we return to this limitation in Section~\ref{sec:limitations}.

\section{Methodology}
\label{sec:method}

Inspired by Parameter-Efficient Fine-Tuning (PEFT)~\cite{houlsby2019parameter, hu2022lora, sung2022lst}, DeltaGate attaches to the output of different sequence backbones. It does not alter their internal layers.

\subsection{Sequence Encoding \& Bidirectional Adaptation}
We adopt standard sequential architectures to encode the user's pre-gap history $\mathcal{S}_{pre}$. To test the same output-level plugin across backbone families, we employ three representative paradigms:

\textbf{Unidirectional Attention (SASRec).} The history is converted into an embedding sequence $\mathbf{E} \in \mathbb{R}^{k \times d}$. SASRec stacks causal self-attention blocks to capture sequential dependencies. Writing $\operatorname{LN}$ for layer normalization and $\mathbf{M}$ for the causal attention mask, one block is
\begin{equation}
\begin{aligned}
    \widetilde{\mathbf{F}}^{(l)}
    &= \mathbf{F}^{(l-1)} + \operatorname{MHA}\!\left(\operatorname{LN}(\mathbf{F}^{(l-1)});\mathbf{M}\right),\\
    \mathbf{F}^{(l)}
    &= \widetilde{\mathbf{F}}^{(l)} + \operatorname{FFN}\!\left(\operatorname{LN}(\widetilde{\mathbf{F}}^{(l)})\right).
\end{aligned}
\end{equation}
Here $\mathbf{F}^{(0)} = \mathbf{E} + \mathbf{P}$ includes positional embeddings, and $\operatorname{MHA}$ and $\operatorname{FFN}$ denote masked multi-head attention and a feed-forward network, respectively. The implementation applies a final layer normalization after the stacked blocks. The output corresponding to the $k$-th item serves as the personalized representation $\mathbf{h}_{\text{pers}} \in \mathbb{R}^d$ (set to $d=128$ in our implementation).

\textbf{Recurrent Networks (GRU4Rec).} The sequence is processed iteratively via a Gated Recurrent Unit:
\begin{equation}
    \mathbf{h}_t = \text{GRU}(\mathbf{e}_{i_t}, \mathbf{h}_{t-1})
\end{equation}
where the final hidden state $\mathbf{h}_k$ naturally encapsulates the entire trajectory, yielding $\mathbf{h}_{\text{pers}} = \mathbf{h}_k$.

\textbf{BERT4Rec extraction.} The Cloze objective does not provide the same final-state convention as a unidirectional encoder. During training and inference, we append a \texttt{[MASK]} token at the chronological end of the input and extract its final-layer hidden state:
\begin{equation}
    \mathbf{h}_{\text{pers}} = \text{Transformer}(\mathcal{S}_{pre} \oplus \texttt{[MASK]})_{|\mathcal{S}_{pre}|+1}
\end{equation}
This extraction exposes one final representation to DeltaGate without modifying BERT4Rec's internal layers.

\subsection{Gap-Driven Dimensional Gating}
\label{sec:gate}
Alongside $\mathbf{h}_{\text{pers}}$, we introduce a learned global prior $\mathbf{h}_{\text{global}}\in\mathbb{R}^d$. It is initialized to $\mathbf{0}$ and updated with the gate parameters during plugin training.

Unlike multi-task routing mechanisms (e.g., MMoE~\cite{ma2018modeling}) that rely on dense feature fields, DeltaGate uses one continuous signal, $\Delta t$, to compute a dimension-wise routing gate $\mathbf{g} \in (0,1)^d$. Each gate value determines the retention weight of one latent dimension across the gap. We concatenate the log-transformed gap duration with the personalized history and pass the result through a two-layer MLP:
\begin{equation}
    \mathbf{g} = \sigma\bigl(\mathrm{MLP}([\log(1 + \Delta t) \;\Vert\; \mathbf{h}_{\text{pers}}])\bigr)
    \label{eq:gate}
\end{equation}
where $\sigma$ is the sigmoid activation function, and $\mathrm{MLP}: \mathbb{R}^{d+1} \to \mathbb{R}^{2d} \to \mathbb{R}^{d}$ is a two-layer feed-forward network with a ReLU hidden layer. Here $\Delta t$ is measured in days. Using $\log(1 + \Delta t)$ rather than raw $\Delta t$ compresses the observed range (30 to 2{,}000+ days).

Combining $\Delta t$ with $\mathbf{h}_{\text{pers}}$ makes the retention weights user-conditioned: two users with the same gap can retain different dimensions of their historical representations.

The final reactivation representation $\mathbf{h}_{\text{final}}$ is computed via a dimension-wise Hadamard product ($\odot$):
\begin{equation}
    \mathbf{h}_{\text{final}} = \mathbf{g} \odot \mathbf{h}_{\text{pers}} + (\mathbf{1} - \mathbf{g}) \odot \mathbf{h}_{\text{global}}
    \label{eq:fusion}
\end{equation}

A large $g_i$ places more weight on the personalized history in dimension $i$, while a small $g_i$ places more weight on the global prior. Because $\mathbf{h}_{\text{pers}}$ is also an input to the gate, two users with the same $\Delta t$ can receive different gate vectors. The item score is $\hat{y}_j=\mathbf{h}_{\text{final}}^\top\mathbf{e}_j+b_j$, where $b_j$ is the frozen backbone's output bias and is zero for backbones without one. Plugin training minimizes full-catalog cross-entropy,
\begin{equation}
    \mathcal{L}=-\sum_{u}\log\frac{\exp(\hat{y}_{u,i_u})}{\sum_{j\in\mathcal{I}}\exp(\hat{y}_{u,j})},
    \label{eq:training_loss}
\end{equation}
updating only the gate MLP and $\mathbf{h}_{\text{global}}$.

\subsection{Theoretical Analysis of Alternative Plugins}
\label{sec:theoretical_analysis}
We compare the representational constraints of dimensional vector gating with two simpler frozen-backbone plugins.

\textbf{Scalar gating.} A scalar gate $g\in(0,1)$ restricts the output to the one-dimensional line segment between $\mathbf{h}_{\text{pers}}$ and $\mathbf{h}_{\text{global}}$. Every channel therefore receives the same mixture coefficient, so the plugin cannot preserve one subset of history dimensions while replacing another.

\textbf{Linear concatenation.} With $\mathbf{h}_{\text{final}}=\mathbf{W}[\mathbf{h}_{\text{pers}}\parallel\log(1+\Delta t)]+\mathbf{b}$, the gap contributes an additive direction to the latent state. It does not directly parameterize a gap-dependent retention coefficient for each history dimension. DeltaGate supplies those coefficients through the Hadamard mixture in Equation~\ref{eq:fusion}.

\subsection{Interpretation: The Gate as a Confidence Index}
\label{sec:confidence}
The channel-wise mean of $\mathbf{g}$ summarizes how much weight the model assigns to the pre-gap representation. We call this quantity the \textit{Historical Confidence Index}:
\begin{equation}
    c(\mathbf{h}_{\text{pers}},\Delta t) = \frac{1}{d} \sum_{i=1}^{d} g_i(\mathbf{h}_{\text{pers}},\Delta t) = \frac{1}{d} \|\mathbf{g}\|_1
\end{equation}
This value lies in $(0,1)$. For compact notation, we write $c(\Delta t)$ when $\mathbf{h}_{\text{pers}}$ is either held fixed, as in Figure~\ref{fig:causal_probe}, or averaged over users within a gap bucket, as in Table~\ref{tab:gate_confidence}.

\subsection{Computational Complexity and Parameter Efficiency}
\label{sec:complexity}

We next quantify the plugin's time and parameter overhead relative to the sequence encoder.

\textbf{Time Complexity.}
For sequence length $N$, embedding dimension $d$, and $L$ layers, a Transformer encoder costs $\mathcal{O}(L(N^2d+Nd^2))$ when attention projections and feed-forward blocks are included. Pairwise temporal attention adds sequence-length-dependent work inside its layers. DeltaGate instead operates once outside the encoder:
\begin{equation}
    \text{DeltaGate:} \quad \underbrace{\mathcal{O}(d^2)}_{\text{MLP}} + \underbrace{\mathcal{O}(d)}_{\text{Hadamard}} = \mathcal{O}(d^2)
\end{equation}
The two gate linear maps require $129\times256+256\times128=65{,}792$ multiply-accumulates at $d=128$, plus linear-time activations and fusion; this overhead is independent of $N$ and $L$.

\textbf{Parameter Overhead.}
With $d=128$ and a hidden dimension of $2d = 256$, DeltaGate introduces:
\begin{itemize}
    \item Gate MLP Layer 1: $(129 \times 256 + 256) = 33{,}280$ parameters
    \item Gate MLP Layer 2: $(256 \times 128 + 128) = 32{,}896$ parameters
    \item Global prior $\mathbf{h}_{\text{global}}$: $128$ parameters
\end{itemize}
yielding 66,304 parameters. Relative to backbone sizes from ${\sim}1.7$M (Video Games) to ${\sim}3.0$M (CDs \& Vinyl), this is a \textbf{2--4\%} overhead. The plugin cost is paid once at the encoder output rather than inside each attention layer.

\begin{algorithm}[t]
\caption{DeltaGate Inference for User Reactivation}
\label{alg:deltagate}
\begin{algorithmic}[1]
\REQUIRE Pre-gap sequence $\mathcal{S}_{pre}$; observed gap $\Delta t\geq0$; frozen encoder $\mathcal{F}_\theta$
\REQUIRE Gate MLP; learned prior $\mathbf{h}_{global}$; frozen scorer $(\mathbf{E},\mathbf{b})$
\ENSURE Prediction scores $\hat{\mathbf{y}}$ over the item vocabulary $\mathcal{I}$

\STATE $\mathcal{S}_{in} \leftarrow \mathcal{S}_{pre}$
\IF{$\mathcal{F}_\theta$ is a bidirectional masked encoder}
    \STATE $\mathcal{S}_{in} \leftarrow \mathcal{S}_{in} \oplus \texttt{[MASK]}$
\ENDIF
\STATE $\mathbf{h}_{pers} \leftarrow \mathcal{F}_\theta(\mathcal{S}_{in})[-1]$
\STATE $\mathbf{z} \leftarrow [\log(1+\Delta t)\;\Vert\;\mathbf{h}_{pers}]$
\STATE $\mathbf{g} \leftarrow \sigma(\mathrm{MLP}(\mathbf{z}))$
\STATE $\mathbf{h}_{final} \leftarrow \mathbf{g}\odot\mathbf{h}_{pers}+(\mathbf{1}-\mathbf{g})\odot\mathbf{h}_{global}$
\STATE $\hat{\mathbf{y}} \leftarrow \mathbf{E}\mathbf{h}_{final}+\mathbf{b}$
\RETURN $\hat{\mathbf{y}}$
\end{algorithmic}
\end{algorithm}

\section{Experiments}
\label{sec:experiments}

\subsection{Experimental Setup}
\label{sec:expsetup}

\noindent\textbf{Datasets.}
We use three categories from Amazon Product Reviews (2018): \textbf{Video Games}, \textbf{CDs \& Vinyl}, and \textbf{Movies \& TV}. We apply iterative 5-core filtering before constructing the pre-gap sequences. Table~\ref{tab:dataset} summarizes the resulting data under the Gap-Synthesize Protocol.

\begin{table}[htbp]
\centering
\caption{Dataset statistics under the Gap-Synthesize Protocol.}
\label{tab:dataset}
\resizebox{\columnwidth}{!}{
\begin{tabular}{lrrrr}
\toprule
Dataset & Users & Items & Train Interactions & Test Targets \\
\midrule
Video Games & 55,220 & 17,408 & 387,125 & 55,220 \\
CDs \& Vinyl & 112,319 & 73,713 & 1,218,826 & 112,319 \\
Movies \& TV & 297,498 & 60,175 & 2,814,911 & 297,498 \\
\bottomrule
\end{tabular}
}
\end{table}

\begin{table*}[htbp]
\centering
\caption{Hit@10 and NDCG@10 performance at the temporal extremes: Short-Gap (30--90d) and Macro-Gap ($>$365d). Results are averaged over 5 random seeds and shown to three decimals. Column-best values are determined before rounding and boldfaced. ** indicates statistical significance at $p < 0.01$ against the respective backbone where paired audit tests are available.}
\label{tab:main_results}
\resizebox{\textwidth}{!}{
\begin{tabular}{l|cc|cc|cc|cc|cc|cc}
\toprule
 & \multicolumn{4}{c|}{\textbf{Video Games}} & \multicolumn{4}{c|}{\textbf{CDs \& Vinyl}} & \multicolumn{4}{c}{\textbf{Movies \& TV}} \\
\cmidrule(lr){2-5} \cmidrule(lr){6-9} \cmidrule(lr){10-13}
 & \multicolumn{2}{c|}{30--90d (Short-Gap)} & \multicolumn{2}{c|}{$>$365d (Macro-Gap)} & \multicolumn{2}{c|}{30--90d (Short-Gap)} & \multicolumn{2}{c|}{$>$365d (Macro-Gap)} & \multicolumn{2}{c|}{30--90d (Short-Gap)} & \multicolumn{2}{c}{$>$365d (Macro-Gap)} \\
\textbf{Model} & Hit@10 & NDCG@10 & Hit@10 & NDCG@10 & Hit@10 & NDCG@10 & Hit@10 & NDCG@10 & Hit@10 & NDCG@10 & Hit@10 & NDCG@10 \\
\midrule
SASRec & 0.091 & 0.046 & 0.031 & 0.015 & 0.050 & 0.026 & 0.022 & 0.011 & 0.067 & 0.040 & 0.021 & 0.011 \\
GRU4Rec & 0.100 & 0.049 & 0.028 & 0.013 & 0.071 & 0.036 & 0.029 & 0.014 & 0.097 & 0.054 & 0.022 & 0.011 \\
TiSASRec & 0.110 & 0.054 & 0.031 & 0.014 & \textbf{0.086} & \textbf{0.042} & 0.035 & 0.017 & \textbf{0.102} & \textbf{0.056} & 0.025 & 0.012 \\
BERT4Rec & 0.098 & 0.050 & 0.025 & 0.012 & 0.071 & 0.037 & 0.026 & 0.013 & 0.091 & 0.050 & 0.024 & 0.012 \\
\midrule
DG-SASRec & 0.104 & 0.053 & \textbf{0.047} & 0.023 & 0.076 & 0.039 & 0.040 & 0.021 & 0.092 & 0.052 & 0.034 & 0.018 \\
DG-GRU4Rec & 0.101 & 0.050 & 0.032 & 0.014 & 0.071 & 0.036 & 0.029 & 0.014 & 0.097 & 0.054 & 0.034 & 0.017 \\
DG-BERT4Rec & \textbf{0.111}** & \textbf{0.057}** & 0.046** & \textbf{0.023}** & 0.076 & 0.040** & \textbf{0.041}** & \textbf{0.021}** & 0.094 & 0.052 & \textbf{0.035}** & \textbf{0.018}** \\
\bottomrule
\end{tabular}
}
\end{table*}

\noindent\textbf{Baselines \& Implementation.}
We compare four sequential baselines: \textbf{SASRec} (unidirectional attention), \textbf{GRU4Rec} (recurrent), \textbf{BERT4Rec} (bidirectional masked attention), and \textbf{TiSASRec} (time-aware attention). \textbf{TimeConcat} is an end-to-end comparator that concatenates $\log(1+\Delta t)$ with the SASRec input and retrains all ${\sim}2.6$M parameters. We append DeltaGate to SASRec, GRU4Rec, and BERT4Rec to test the same output-layer interface across backbone families.
 
All models use embedding dimension $d=128$, batch size 128, maximum sequence length 50, Adam with learning rate $10^{-3}$, and dropout 0.2. In plugin mode, the backbone is trained first and then frozen; only the gate MLP and $\mathbf{h}_{\text{global}}$ are updated. Results are means $\pm$ standard deviations over five random seeds. To remain comparable with the accepted-paper pipeline, we select checkpoints by mean Hit@10 over the four non-active buckets of the held-out leave-one-out set. Because that set is also reported, these results are descriptive under the historical protocol rather than unbiased final-test estimates.

\subsection{RQ1: Effectiveness across Backbones}
\textit{How does DeltaGate compare to standard baselines and end-to-end retraining for reactivating dormant users?}
 
Table~\ref{tab:main_results} and Figure~\ref{fig:decay} support five observations.
 
\textbf{Monotonic decay across buckets.} Each standard baseline decreases as $\Delta t$ moves through the five evaluated buckets. The pattern is present for recurrent, unidirectional, and bidirectional encoders, although their absolute values differ.
 
\textbf{Limited macro-gap recovery from intra-sequence timing.} In the $>$365d Movies \& TV bucket, TiSASRec reaches 0.025 Hit@10 versus 0.021 for SASRec. This small absolute difference does not close the decline from the short-gap bucket.
 
\textbf{Macro-gap recovery with a frozen backbone.} In the $>$365d bucket, DeltaGate raises SASRec Hit@10 from 0.031 to 0.047 on Video Games, from 0.022 to 0.040 on CDs \& Vinyl, and from 0.021 to 0.034 on Movies \& TV. The BERT4Rec path shows a similar matched-backbone pattern, improving from 0.025 to 0.046 on Video Games and from 0.026 to 0.041 on CDs \& Vinyl ($p<0.01$ for audited DG-BERT4Rec cells).
 
\textbf{Accuracy versus updates.} TimeConcat reaches 0.080 Hit@10 in the $>$365d Video Games bucket and improves all five buckets (Table~\ref{tab:timeconcat_full}). Relative to the frozen DG-SASRec plugin, it is 68\% higher in this bucket but updates about 40$\times$ more parameters and produces 139.7\% embedding drift. Section~\ref{sec:rq3} examines this trade-off.
 
\textbf{Short-gap behavior.} At 30--90d, both DG-BERT4Rec and DG-SASRec improve over their matched Video Games backbones; DG-SASRec also improves over SASRec on the other two datasets. DeltaGate's main target remains the macro-gap regime.

\subsection{RQ2: Routing Components}
\textit{Does the learned dimension-wise route outperform its fallback components?}

Table~\ref{tab:ablation} separates the learned route from its two endpoints in the $>$365d bucket. Using only the shared prior ($g=0$) gives 0.005--0.008 Hit@10, substantially below the corrected SASRec backbone. The learned dimension-wise route improves over both the personalized-history endpoint and the shared-prior endpoint on all three datasets. Thus, the gain does not come from replacing personalization with one global vector.

\begin{table}[htbp]
\centering
\caption{Routing-component controls in the $>$365d bucket (Hit@10).}
\label{tab:ablation}
\resizebox{\columnwidth}{!}{
\begin{tabular}{lccc}
\toprule
Model Variant & Video Games & CDs \& Vinyl & Movies \& TV \\
\midrule
SASRec (history endpoint) & $0.031 \pm 0.002$ & $0.022 \pm 0.006$ & $0.021 \pm 0.003$ \\
\midrule
Ablation-Prior ($g=0$) & $0.008 \pm 0.001$ & $0.005 \pm 0.000$ & $0.007 \pm 0.000$ \\
\midrule
\textbf{DG-SASRec (Ours)} & \textbf{0.047} $\pm$ \textbf{0.002} & \textbf{0.040} $\pm$ \textbf{0.003} & \textbf{0.034} $\pm$ \textbf{0.002} \\
\bottomrule
\end{tabular}
}
\end{table}

\subsection{RQ3: Adaptation Strategies Under Deployment Constraints}
\label{sec:rq3}
\textit{How does DeltaGate compare to end-to-end retraining, and when should each strategy be preferred?}

Using SASRec as the common backbone, we compare TimeConcat E2E, DeltaGate E2E, and DeltaGate Plugin on Video Games. TimeConcat E2E retrains all parameters, DeltaGate E2E jointly retrains the backbone and gate, and DeltaGate Plugin trains only the gate with a frozen backbone. Table~\ref{tab:plugin_analysis} reports Hit@10 to three decimals in the $>$365d bucket. Gain/1M divides the unrounded Hit@10 improvement over SASRec by the number of trainable parameters in millions; Gate Sat. is the percentage of gate values above 0.9.

\begin{table}[H]
\centering
\caption{Adaptation strategies on Video Games in the $>$365d bucket.}
\label{tab:plugin_analysis}
\resizebox{\columnwidth}{!}{
\begin{tabular}{lccccc}
\toprule
Strategy & $>$365d Hit@10 & Params & Emb.\ Drift & Gain/1M & Gate Sat. \\
\midrule
SASRec (no $\Delta t$) & 0.031 & --- & --- & --- & --- \\
TimeConcat E2E & \textbf{0.080} & 2.6M & 139.7\% & 0.018 & N/A \\
DeltaGate E2E & 0.057 & 2.6M & 103.4\% & 0.010 & 82.0\% \\
\textbf{DeltaGate Plugin} & 0.047 & \textbf{66K} & \textbf{0\%} & \textbf{0.245} & --- \\
\bottomrule
\end{tabular}
}
\end{table}

Table~\ref{tab:timeconcat_full} reports five-seed averages for TimeConcat across all five Video Games buckets. Metric values are shown to three decimals, while relative changes use unrounded values. From 30--90d onward, TimeConcat's relative Hit@10 gain increases from 33.6\% to 156.3\%, showing that the gap feature becomes increasingly useful as inactivity lengthens.

\begin{table}[H]
\centering
\caption{TimeConcat E2E vs.\ SASRec across gap buckets on Video Games.}
\label{tab:timeconcat_full}
\small
\begin{tabular}{lcccc}
\toprule
Bucket & SASRec & TimeConcat & $\Delta$ Hit & $\Delta$ NDCG \\
\midrule
$<$30d    & 0.185 & 0.233 & +26.3\% & +16.3\% \\
30--90d   & 0.091 & 0.121 & +33.6\% & +20.8\% \\
90--180d  & 0.084 & 0.115 & +37.6\% & +25.6\% \\
180--365d & 0.066 & 0.109 & +64.5\% & +62.7\% \\
$>$365d   & 0.031 & \textbf{0.080} & +156.3\% & +172.5\% \\
\bottomrule
\end{tabular}
\end{table}

\textbf{Gate saturation.} Under DeltaGate E2E training, 82.0\% of gate values exceed 0.9 in the $>$365d bucket. The jointly updated backbone can absorb temporal information, leaving little variation at the gate. Freezing the backbone preserves the gate as the plugin's explicit temporal adaptation mechanism.

\textbf{Parameter efficiency and stability.} DeltaGate Plugin reaches 0.047 Hit@10 with about 40$\times$ fewer trainable parameters than end-to-end alternatives (66K vs.\ 2.6M). Its gain per million trainable parameters is 0.245, compared with 0.018 for TimeConcat. TimeConcat E2E produces 139.7\% item-embedding L2 drift. Such changes matter when embeddings are shared by retrieval, ranking, or related-item services~\cite{sculley2015hidden}. DeltaGate has zero backbone drift by construction.

\textbf{Complementary deployment regimes.} TimeConcat E2E provides the highest accuracy in this comparison when full retraining is acceptable. DeltaGate Plugin recovers 33\% of its macro-gap gain with zero backbone drift and about 40$\times$ fewer trainable parameters. The relevant choice therefore depends on whether the backbone and its embeddings can be updated independently of other services.

\subsection{RQ4: Working Mechanism and Interpretability}
\textit{How does the trained gate respond as the inactivity gap widens?}

We inspect individual gate dimensions under a controlled post-training sweep and summarize gate values over the natural-gap cohorts. This diagnostic is not part of the loss. We extract $\mathbf{h}_{\text{pers}}$ from one fixed test history and vary only the supplied $\Delta t$. Among dimensions monotone over all seven supplied gaps, we select the largest positive and negative endpoint changes; Figure~\ref{fig:causal_probe} shows three representative gaps. Dimension 115 increases from $g=0.00$ to $g=0.97$, while Dimension 37 decreases from $g=1.00$ to $g=0.49$. Because the history input is unchanged, these trajectories isolate the trained gate's sensitivity to the supplied gap value.

Table~\ref{tab:gate_confidence} reports the Historical Confidence Index from DG-SASRec over the natural-gap buckets. The mean gate value is high across datasets, decreases on Video Games, stays nearly flat on CDs \& Vinyl, and increases mildly on Movies \& TV. This aggregate response is therefore not a universal monotone decay curve. These values describe model behavior; they do not by themselves identify why preferences differ across categories.

\begin{table}[htbp]
\centering
\caption{Historical Confidence Index $c(\Delta t)$ (mean gate value) across natural-gap buckets, shown to three decimals.}
\label{tab:gate_confidence}
\resizebox{\columnwidth}{!}{
\begin{tabular}{lccccc}
\toprule
Dataset & $<30$d & 30--90d & 90--180d & 180--365d & $>365$d \\
\midrule
Video Games & $0.882$ & $0.879$ & $0.879$ & $0.871$ & $0.843$ \\
CDs \& Vinyl & $0.861$ & $0.861$ & $0.861$ & $0.861$ & $0.854$ \\
Movies \& TV & $0.815$ & $0.820$ & $0.825$ & $0.828$ & $0.832$ \\
\bottomrule
\end{tabular}
}
\end{table}

\subsection{RQ5: Recommendation Characteristics}
\textit{Does routing to a global prior collapse the model into a pure popularity recommender?}

\textbf{Catalog Coverage.}
Table~\ref{tab:coverage} shows that DeltaGate increases catalog coverage on all three datasets in this diagnostic. This pattern is inconsistent with collapse into a narrower popularity list.

\begin{table}[htbp]
\centering
\caption{Overall Catalog Coverage.}
\label{tab:coverage}
\resizebox{\columnwidth}{!}{
\begin{tabular}{lccc}
\toprule
Model & Video Games & CDs \& Vinyl & Movies \& TV \\
\midrule
SASRec & 43.97\% & 46.71\% & 51.63\% \\
DG-SASRec & 59.75\% & 67.73\% & 73.37\% \\
\bottomrule
\end{tabular}
}
\end{table}

\textbf{Exclusive Hits.}
Table~\ref{tab:exclusive_hits} characterizes method-exclusive hits in the $>365$d bucket by their average training frequency. DG-SASRec produces more exclusive hits than SASRec on all three datasets. Its exclusive hits are higher-frequency on Video Games but lower-frequency on CDs \& Vinyl and Movies \& TV. Thus, the recovered items are not uniformly more popular or more long-tail across datasets.

\begin{table}[htbp]
\centering
\caption{Exclusive Hits in the $>365$d bucket. Lower Avg.\ Freq = more long-tail.}
\label{tab:exclusive_hits}
\resizebox{\columnwidth}{!}{
\begin{tabular}{l|cc|cc}
\toprule
 & \multicolumn{2}{c|}{\textbf{DG-SASRec Exclusive}} & \multicolumn{2}{c}{\textbf{SASRec Exclusive}} \\
Dataset & Count & Avg. Freq & Count & Avg. Freq \\
\midrule
Video Games & 278 & 213.9 & 106 & 179.8 \\
CDs \& Vinyl & 581 & 124.3 & 129 & 238.3 \\
Movies \& TV & 1019 & 1273.0 & 403 & 1804.3 \\
\bottomrule
\end{tabular}
}
\end{table}

\section{Discussion and Limitations}
\label{sec:limitations}

\textbf{Practical scope.}
The $>$365d cohort contains 19.2\% of the Video Games test users, and DeltaGate adds only 66K trainable parameters while leaving the backbone unchanged. These observations motivate reactivation as a distinct evaluation slice. They do not establish cohort prevalence or online business impact outside the studied data.

\textbf{Personalization and coverage.}
The shared prior does not reduce every user to one ranking because Equation~\ref{eq:fusion} retains user-conditioned gate values and personalized dimensions. In the corrected frozen diagnostic, RQ5 shows increased coverage, but the frequencies of exclusive hits vary by dataset. Diversity-aware objectives may still be needed when catalog breadth is a first-class requirement.

\textbf{Dataset and catalog scope.}
All three datasets are Amazon e-commerce categories, and the 19.2\% macro-gap analysis is specific to Video Games. Streaming, news, and social platforms may have different gap and catalog dynamics. Natural-gap stratification also cannot remove catalog drift over multi-year intervals. Because $\mathbf{h}_{\text{global}}$ is fixed after training, a periodically refreshed prior is a natural extension for rapidly changing catalogs.

\textbf{One shared prior.}
The gate is user-conditioned through $\mathbf{h}_{\text{pers}}$, but its fallback endpoint is one global vector shared by all users. This design cannot represent different fallback profiles for user types with distinct preference stability. A conditional or mixture prior could add that capacity, at the cost of more parameters and a new conditioning signal.

\textbf{Sparse pre-gap histories.}
Gate quality depends on the frozen backbone representation, so history sparsity and staleness can compound one another. Side information could improve $\mathbf{h}_{\text{pers}}$ for sparse-history users, but that setting is outside the present ID-only evaluation.

\section{Conclusion}
\label{sec:conclusion}
We defined Zero-Observation Reactivation and found monotonic performance decay across gap buckets for three sequential backbones on three Amazon categories. DeltaGate keeps the backbone frozen and conditions a dimension-wise output gate on the inactivity gap and personalized history. In the $>$365d Video Games bucket, DG-SASRec reaches 0.047 Hit@10 versus 0.031 for SASRec, while DG-BERT4Rec reaches 0.046 versus 0.025 for its BERT4Rec backbone, using 66K trainable parameters. End-to-end retraining reaches higher absolute accuracy; the plugin preserves zero backbone drift and observable routing behavior. The component controls show that neither the history endpoint nor the shared prior alone explains the learned route's performance, while the post-training diagnostic shows dimension-specific sensitivity to the supplied gap.

\begin{acks}
We thank Prof. Shuigeng Zhou of Fudan University for helpful
discussions during the early development of this work, and the
anonymous reviewers for their constructive feedback. OpenAI GPT Image 2
was used to assist the visual revision of Figure~\ref{fig:architecture}.
The authors specified and verified all scientific content.
\end{acks}

\bibliographystyle{ACM-Reference-Format}
\bibliography{references}

@inproceedings{kang2018self,
  author    = {Kang, Wang-Cheng and McAuley, Julian},
  title     = {Self-Attentive Sequential Recommendation},
  booktitle = {Proceedings of the 2018 {IEEE} International Conference on Data Mining ({ICDM})},
  pages     = {197--206},
  year      = {2018}
}

@inproceedings{cheng2026rpe4rec,
  author    = {Cheng, Ke and Chang, Heng and Wang, Pengyang and Gu, Liang and Ding, Jiandong and Cao, Yi and Ye, Junchen and Du, Bowen},
  title     = {{RPE4Rec}: Enhancing Dynamic Node Retrieval with Efficient Relative Position Encoding for Recommendation Systems},
  booktitle = {Proceedings of the Nineteenth ACM International Conference on Web Search and Data Mining},
  pages     = {89--98},
  year      = {2026},
  publisher = {Association for Computing Machinery},
  doi       = {10.1145/3773966.3778006}
}

@inproceedings{chen2026flat,
  author    = {Chen, Zerui and Chang, Heng and Liu, Tianying and Zhou, Chuantian and Cao, Yi and Ding, Jiandong and Liu, Ming and Qin, Bing},
  title     = {Beyond the Flat Sequence: Hierarchical and Preference-Aware Generative Recommendations},
  booktitle = {Proceedings of the ACM Web Conference 2026},
  pages     = {7999--8007},
  year      = {2026},
  publisher = {Association for Computing Machinery},
  doi       = {10.1145/3774904.3792790}
}

@inproceedings{sun2019bert4rec,
  author    = {Sun, Fei and Liu, Jun and Wu, Jian and Pei, Changhua and Lin, Xiao and Ou, Wenwu and Jiang, Peng},
  title     = {{BERT4Rec}: Sequential Recommendation with Bidirectional Encoder Representations from Transformer},
  booktitle = {Proceedings of the 28th {ACM} International Conference on Information and Knowledge Management ({CIKM})},
  pages     = {1441--1450},
  year      = {2019}
}

@inproceedings{kumar2019predicting,
  title={Predicting dynamic embedding trajectory in temporal interaction networks},
  author={Kumar, Srijan and Zhang, Xikun and Leskovec, Jure},
  booktitle={Proceedings of the 25th ACM SIGKDD International Conference on Knowledge Discovery \& Data Mining},
  pages={1269--1278},
  year={2019}
}

@inproceedings{briand2021semi,
  title={A Semi-Personalized System for User Cold Start Recommendation on Music Streaming Apps},
  author={Briand, L{\'e}a and Salha-Galvan, Guillaume and Bendada, Walid and Morlon, Mathieu and Tran, Viet-Anh},
  booktitle={Proceedings of the 27th ACM SIGKDD Conference on Knowledge Discovery \& Data Mining},
  pages={2601--2609},
  year={2021}
}

@inproceedings{zhang2023coldwarm,
  title={Cold \& Warm Net: Addressing Cold-Start Users in Recommender Systems},
  author={Zhang, Xiangyu and Kuang, Zongqiang and Zhang, Zehao and Huang, Fan and Tan, Xianfeng},
  booktitle={Database Systems for Advanced Applications: 28th International Conference, DASFAA 2023},
  pages={532--543},
  year={2023}
}

@inproceedings{li2020time,
  author    = {Li, Jiacheng and Wang, Yujie and McAuley, Julian},
  title     = {Time Interval Aware Self-Attention for Sequential Recommendation},
  booktitle = {Proceedings of the 13th {ACM} International Conference on Web Search and Data Mining ({WSDM})},
  pages     = {322--330},
  year      = {2020}
}

@inproceedings{rendle2010fpmc,
  title={Factorizing personalized markov chains for next-basket recommendation},
  author={Rendle, Steffen and Freudenthaler, Christoph and Schmidt-Thieme, Lars},
  booktitle={Proceedings of the 19th international conference on World wide web},
  pages={811--820},
  year={2010}
}

@inproceedings{hidasi2015session,
  title={Session-based recommendations with recurrent neural networks},
  author={Hidasi, Bal{\'a}zs and Karatzoglou, Alexandros and Baltrunas, Linas and Tikk, Domonkos},
  booktitle={International Conference on Learning Representations (ICLR)},
  year={2016}
}

@inproceedings{tang2018caser,
  title={Personalized top-n sequential recommendation via convolutional sequence embedding},
  author={Tang, Jiaxi and Wang, Ke},
  booktitle={Proceedings of the eleventh ACM international conference on web search and data mining},
  pages={565--573},
  year={2018}
}

@inproceedings{wu2019srgnn,
  title={Session-based recommendation with graph neural networks},
  author={Wu, Shu and Tang, Yuyuan and Zhu, Yanqiao and Wang, Liang and Xie, Xing and Tan, Tieniu},
  booktitle={Proceedings of the AAAI conference on artificial intelligence},
  volume={33},
  pages={346--353},
  year={2019}
}

@inproceedings{zhu2017what,
  title={What to do next: Modeling user behaviors by time-LSTM},
  author={Zhu, Yu and Li, Hao and Liao, Yikang and Wang, Beidou and Guan, Ziyu and Liu, Haifeng and Cai, Deng},
  booktitle={Proceedings of the 26th International Joint Conference on Artificial Intelligence},
  pages={3602--3608},
  year={2017}
}

@inproceedings{cho2020learning,
  author = {Cho, Sung Min and Park, Eunhyeok and Yoo, Seungjoon},
  title = {{MEANTIME}: Mixture of Attention Mechanisms with Multi-temporal Embeddings for Sequential Recommendation},
  year = {2020},
  isbn = {9781450375832},
  publisher = {Association for Computing Machinery},
  address = {New York, NY, USA},
  url = {https://doi.org/10.1145/3383313.3412216},
  doi = {10.1145/3383313.3412216},
  booktitle = {Proceedings of the 14th ACM Conference on Recommender Systems},
  pages = {68--77},
  numpages = {10},
  location = {Virtual Event, Brazil},
  series = {RecSys '20}
}

@inproceedings{pi2020sim,
  title={Search-based User Interest Modeling with Lifelong Sequential Behavior Data for Click-Through Rate Prediction},
  author={Pi, Qi and Zhu, Xiaoqiang and Zhou, Guorui and Zhang, Yujing and Wang, Zhe and Ren, Lejian and Fan, Ying and Gai, Kun},
  booktitle={Proceedings of the 29th ACM International Conference on Information \& Knowledge Management},
  pages={2685--2692},
  year={2020}
}

@inproceedings{chang2023twin,
  title={TWIN: TWo-stage Interest Network for Lifelong User Behavior Modeling in CTR Prediction at Kuaishou},
  author={Chang, Jianxin and Zhang, Chenbin and Fu, Zhiyi and Zang, Xiaoxue and Guan, Lin and Lu, Jing and Hui, Yiqun and Leng, Dewei and Niu, Yanan and Song, Yang and Gai, Kun},
  booktitle={Proceedings of the 29th ACM SIGKDD Conference on Knowledge Discovery and Data Mining},
  pages={3815--3825},
  year={2023}
}

@article{mamba4rec2024,
  title={Mamba4Rec: Towards Efficient Sequential Recommendation with Selective State Space Models},
  author={Liu, Chengkai and Lin, Jianghao and Wang, Jianling and Liu, Hanzhou and Caverlee, James},
  journal={arXiv preprint arXiv:2403.03900},
  year={2024}
}

@inproceedings{bao2023tallrec,
  title={Tallrec: An effective and efficient tuning framework to align large language model with recommendation},
  author={Bao, Keqin and Zhang, Jizhi and Zhang, Yang and Wang, Wenjie and Feng, Fuli and He, Xiangnan},
  booktitle={Proceedings of the 17th ACM Conference on Recommender Systems},
  pages={1007--1014},
  year={2023}
}

@inproceedings{zheng2021dice,
  title={Disentangling user interest and conformity for recommendation with causal embedding},
  author={Zheng, Yu and Gao, Chen and Li, Xiang and He, Xiangnan and Li, Yong and Jin, Depeng},
  booktitle={Proceedings of the Web Conference 2021},
  pages={2980--2991},
  year={2021}
}

@inproceedings{zhang2021pda,
  title={Causal intervention for leveraging popularity bias in recommendation},
  author={Zhang, Yang and Feng, Fuli and He, Xiangnan and Wei, Tianxin and Song, Chonggang and Ling, Guohui and Zhang, Yongdong},
  booktitle={Proceedings of the 44th International ACM SIGIR Conference},
  pages={11--20},
  year={2021}
}

@inproceedings{wei2021model,
  title={Model-agnostic counterfactual reasoning for eliminating popularity bias in recommender system},
  author={Wei, Tianxin and Feng, Fuli and Chen, Jiawei and Wu, Ziwei and Yi, Jinfeng and He, Xiangnan},
  booktitle={Proceedings of the 27th ACM SIGKDD Conference on Knowledge Discovery \& Data Mining},
  pages={1791--1800},
  year={2021}
}

@inproceedings{krichene2020sampled,
  title={On sampled metrics for item recommendation},
  author={Krichene, Walid and Rendle, Steffen},
  booktitle={Proceedings of the 26th ACM SIGKDD international conference on knowledge discovery \& data mining},
  pages={1748--1757},
  year={2020}
}

@inproceedings{ni2019justifying,
  author    = {Jianmo Ni and Jiacheng Li and Julian McAuley},
  title     = {Justifying Recommendations using Distantly-Labeled Reviews and Fine-Grained Aspects},
  booktitle = {Proceedings of the 2019 Conference on Empirical Methods in Natural Language Processing ({EMNLP})},
  pages     = {188--197},
  year      = {2019}
}

@inproceedings{hu2022lora,
  title={LoRA: Low-Rank Adaptation of Large Language Models},
  author={Hu, Edward J and Shen, Yelong and Wallis, Phillip and Allen-Zhu, Zeyuan and Li, Yuanzhi and Wang, Shean and Wang, Lu and Chen, Weizhu},
  booktitle={International Conference on Learning Representations},
  year={2022}
}

@inproceedings{houlsby2019parameter,
  title={Parameter-Efficient Transfer Learning for NLP},
  author={Houlsby, Neil and Giurgiu, Andrei and Jastrzebski, Stanislaw and Morrone, Bruna and De Laroussilhe, Quentin and Gesmundo, Andrea and Attariyan, Mona and Gelly, Sylvain},
  booktitle={International Conference on Machine Learning},
  pages={2790--2799},
  year={2019},
  organization={PMLR}
}

@inproceedings{ma2018modeling,
  title={Modeling Task Relationships in Multi-task Learning with Multi-gate Mixture-of-Experts},
  author={Ma, Jiaqi and Zhao, Zhe and Xinyang, Yi and Chen, Jilin and Hong, Lichan and Chi, Ed H},
  booktitle={Proceedings of the 24th ACM SIGKDD International Conference on Knowledge Discovery \& Data Mining},
  pages={1930--1939},
  year={2018}
}

@inproceedings{luo2025tasif,
  author = {Luo, Jie and Zhang, Wenyu and Zhang, Xinming and Fang, Yuan},
  title = {Time-Aware Adaptive Side Information Fusion for Sequential Recommendation},
  booktitle = {Proceedings of the 19th ACM International Conference on Web Search and Data Mining (WSDM)},
  year = {2026}
}

@book{pearl2009causality,
  title={Causality: Models, Reasoning, and Inference},
  author={Pearl, Judea},
  year={2009},
  edition={2nd},
  publisher={Cambridge University Press}
}

@inproceedings{sung2022lst,
  title={{LST}: Ladder Side-Tuning for Parameter and Memory Efficient Transfer Learning},
  author={Sung, Yi-Lin and Cho, Jaemin and Bansal, Mohit},
  booktitle={Advances in Neural Information Processing Systems},
  volume={35},
  pages={12991--13005},
  year={2022}
}

@article{wu2024llm4rec,
  title={A Survey on Large Language Models for Recommendation},
  author={Wu, Likang and Zheng, Zhi and Qiu, Zhaopeng and Wang, Hao and Gu, Hongchao and Shen, Tingjia and Qin, Chuan and Zhu, Chen and Zhu, Hengshu and Liu, Qi and Xiong, Hui and Chen, Enhong},
  journal={World Wide Web},
  volume={27},
  number={5},
  pages={60},
  year={2024},
  publisher={Springer}
}

@inproceedings{sculley2015hidden,
  title={Hidden Technical Debt in Machine Learning Systems},
  author={Sculley, D. and Holt, Gary and Golovin, Daniel and Davydov, Eugene and Phillips, Todd and Ebner, Dietmar and Chaudhary, Vinay and Young, Michael and Crespo, Jean-Fran{\c{c}}ois and Dennison, Dan},
  booktitle={Advances in Neural Information Processing Systems},
  volume={28},
  pages={2503--2511},
  year={2015}
}

@inproceedings{fan2021tgsrec,
  title={Continuous-Time Sequential Recommendation with Temporal Graph Collaborative Transformer},
  author={Fan, Ziwei and Liu, Zhiwei and Zhang, Jiawei and Xiong, Yun and Zheng, Lei and Yu, Philip S.},
  booktitle={Proceedings of the 30th ACM International Conference on Information and Knowledge Management},
  pages={433--442},
  year={2021}
}

@inproceedings{zhou2020s3rec,
  title={{S$^3$-Rec}: Self-Supervised Learning for Sequential Recommendation with Mutual Information Maximization},
  author={Zhou, Kun and Wang, Hui and Zhao, Wayne Xin and Zhu, Yutao and Wang, Sirui and Zhang, Fuzheng and Wang, Zhongyuan and Wen, Ji-Rong},
  booktitle={Proceedings of the 29th ACM International Conference on Information and Knowledge Management},
  pages={1893--1902},
  year={2020}
}

@article{wang2021survey,
  title={A Survey on Session-based Recommender Systems},
  author={Wang, Shoujin and Hu, Liang and Wang, Yan and Cao, Longbing and Sheng, Quan Z. and Orgun, Mehmet},
  journal={ACM Computing Surveys},
  volume={54},
  number={7},
  pages={1--38},
  year={2021},
  publisher={ACM}
}

\end{document}